\documentclass[10pt, conference, letterpaper]{IEEEtran}
\IEEEoverridecommandlockouts
\usepackage{cite}
\usepackage{amssymb,amsmath,amsthm,amsfonts}
\usepackage{graphicx}
\usepackage{textcomp}
\usepackage{xcolor}
\usepackage[hyphens]{url}
\usepackage{hyperref}
\usepackage{float}
\usepackage[linesnumbered,ruled, boxed, lined, noend]{algorithm2e}
\usepackage{algpseudocode}
\usepackage{cancel}
\usepackage{comment}
\def\BibTeX{{\rm B\kern-.05em{\sc i\kern-.025em b}\kern-.08em
    T\kern-.1667em\lower.7ex\hbox{E}\kern-.125emX}}

\DeclareMathOperator*{\argmin}{argmin}

\newcommand{\Cb}[1]{ \mathrm{C}_{#1} }

\begin{document}

\title{Hierarchical Edge-Cloud Task Offloading in NTN for Remote Healthcare \\
}

\author{

\IEEEauthorblockN{ Alejandro Flores$^*$; Danial Shafaie$^+$, Konstantinos Ntontin$^*$, Elli Kartsakli$^+$, Symeon Chatzinotas$^*$ }
\IEEEauthorblockA{$^*$Interdisciplinary Centre for Security, Reliability and Trust (SnT), University of Luxembourg, Luxembourg.}
\IEEEauthorblockA{$^+$Barcelona Supercomputing Center, Spain.}
\IEEEauthorblockA{e-mails:$^*$\{alejandro.flores, kostantinos.ntontin, symeon.chatzinotas\}@uni.lu;}
\IEEEauthorblockA{ $^+$\{danial.shafaie, elli.kartsakli\}@bsc.es}\vspace{-2em}
}

\newcommand{\SAT}{\mathrm{SAT}}
\newcommand{\LEO}{\mathrm{LEO}}
\newcommand{\UAV}{\mathrm{UAV}}
\newcommand{\HAPS}{\mathrm{HAPS}}
\newcommand{\IoT}{\mathrm{IoT}}
\newcommand{\maxt}{\mathrm{max}}
\newcommand{\mint}{\mathrm{min}}

\maketitle

\begin{abstract}
In this work, we study a hierarchical non-terrestrial network as an edge-cloud platform for remote computing of tasks generated by remote ad-hoc healthcare facility deployments, or internet of medical things (IoMT) devices. We consider a high altitude platform station (HAPS) to provide local multiaccess edge server (MEC) services to a set of remote ground medical devices, and a low-earth orbit (LEO) satellite, serving as a bridge to a remote cloud computing server through a ground gateway (GW), providing a large amount of computing resources to the HAPS. In this hierarchical system, the HAPS and the cloud server charges the ground users and the HAPS for the use of the spectrum and the computing of their tasks respectively. Each tier seeks to maximize their own utility in a selfish manner. To encourage the prompt computation of the tasks, a local delay cost is assumed. We formulate the optimal per-task cost at each tier that influences the corresponding offloading policies, and find the corresponding optimal bandwidth allocation.
\end{abstract}

\begin{IEEEkeywords}
multi-access edge computing (MEC), non-terrestrial networks, resource allocation, task offloading, healthcare 4.0.
\end{IEEEkeywords}

% \section{Introduction}\label{sec:Introduction}

\section{Introduction}\label{sec:Motivation}

% \section{Motivation and Social Impact}\label{sec:Motivation}

Healthcare 4.0 (H4.0) envisions the persistent data gathering, sharing and processing from patients to healthcare providers and healthcare entities, which enables enhanced treatment through remote monitoring, proactive treatment, disease prevention and intelligent disease treatment~\cite{art:Healthcare40}. Biosignal data gathered from patients, through specialized healthcare sensors, or through internet of medical things (IoMT) devices enables making appropriate health-related inferences~\cite{art:IoMT_fever,art:IoMT_cough}. 
% ~\cite{art:IoMT_fever,art:IoMT_cough,art:IoMT_fatigue}. 
% One element of H4.0 is the use of internet of medical things (IoMT) devices, that monitor biosignals of patients and are able to make appropriate inferences with the gathered data~\cite{art:IoMT_fever,art:IoMT_cough,art:IoMT_fatigue}. 
These inferences are abstracted as computing tasks that can be processed locally, or offloaded to remote edge or cloud servers if the local resources are constrained, or if the results need to be allocated particularly promptly.
% for a prompt computation.
% This data needs to be deployed to remote healthcare facilities for reporting and analysis. 
% However, an IoMT wearable user may lie outside the range of ground cellular networks to utilize to offload their data, given that only 15\% of the surface of the Earth has cellular coverage~\cite{web:wef2024}. 
% However, i
In remote locations, disaster situations~\cite{web:DisasterHC}, or war zones~\cite{web:WarHC}, ad-hoc healthcare facilities can be deployed to provide healthcare services to people in need. However, due to the nature of these environments, a local ground cellular network may not be available due to their destruction, congestion, or remoteness of the area, given that only 15\% of the surface of the Earth has cellular coverage~\cite{web:wef2024}. 
% However, patients or users may lie outside the range of ground cellular networks to offload their data.

As a solution to these problems, non-terrestrial networks (NTNs), which comprise air nodes such as high altitude platform stations (HAPS), and space nodes such as low-earth orbit (LEO) satellites, can act as multi-access edge servers (MEC) or bridges to remote mobile cloud computing (MCC) servers, to offload computing tasks generated by the ground healthcare nodes. Connection to satellite constellations orbiting the Earth, in general belonging to different companies~\cite{web:NumberOfSats}, is generally performed by acquiring the corresponding user terminal and subscription from the vendor. 
% Although it is envisioned that LEO satellites will work in regenerative-payload mode, currently 
LEO satellites are typically resource-constrained and better suited to relay information to a remote cloud server within coverage~\cite{art:transparentpayload}. HAPS may be deployed on a case-by-case basis to directed zones, being able to connect to arbitrary user terminals~\cite{gsma2022haps} and accommodate larger payloads, allowing greater computing capacity. 
There is recent literature on the use of hybrid TN-NTNs for task offloading. In~\cite{FloresCabezas2025} we explored the non-orthogonal resource initialization of NTN MEC-enabled nodes for task offloading to a LEO satellite through intermediate UAVs. Hybrid MEC-enabled satellite networks have also been studied, such as in~\cite{10713345} via a two-level hierarchical game for edge–cloud collaboration and joint task offloading and resource allocation. Pricing and hierarchy also appear in~\cite{9289259} which studies a Stackelberg game in TN–NTN relaying and in robustness settings such as in~\cite{11174369}. Also, a complete model for selfish task offloading was studied in~\cite{Shafaie2025}, introducing a Game theoretic modeling. 
% for anti-jamming applications.
% Focusing on offloading,~\cite{10829804} studies computation offloading by MEC servers, where devices select processing under limited communication time and shared resources.
% the following paragraph serves as a step for our work, so we will change it together
% check my changes in equations with Alejandro
% Why the bandwidth cost for HAPS is unavoidable by not using the infrastructure? like the GD
Targeting the research gaps, our work studies remote healthcare applications in NTN with a three-tier non-collaborative computing infrastructure. Unlike prior studies that mainly optimize offloading and computing resource allocation, we derive the optimal bandwidth allocation scheme, while proposing a game theoretic solution for offloading decisions and setting prices, for healthcare-related data models.

% Expanding upon the literature, in this work we propose a NTN hierarchical architecture in which healthcare ground devices can compute their generated tasks, or pay for the bandwidth usage to access a local MEC-enabled HAPS for swifter task computation. The HAPS can choose to compute the tasks locally, or pay for the bandwidth to use a LEO satellite within range, to relay the tasks to a remote cloud server for the fastest computing. 

\section{System Model}\label{sec:Description}

% \subsection{System Model}

\begin{figure}[ht]
    \centering
    \includegraphics[width=1\linewidth]{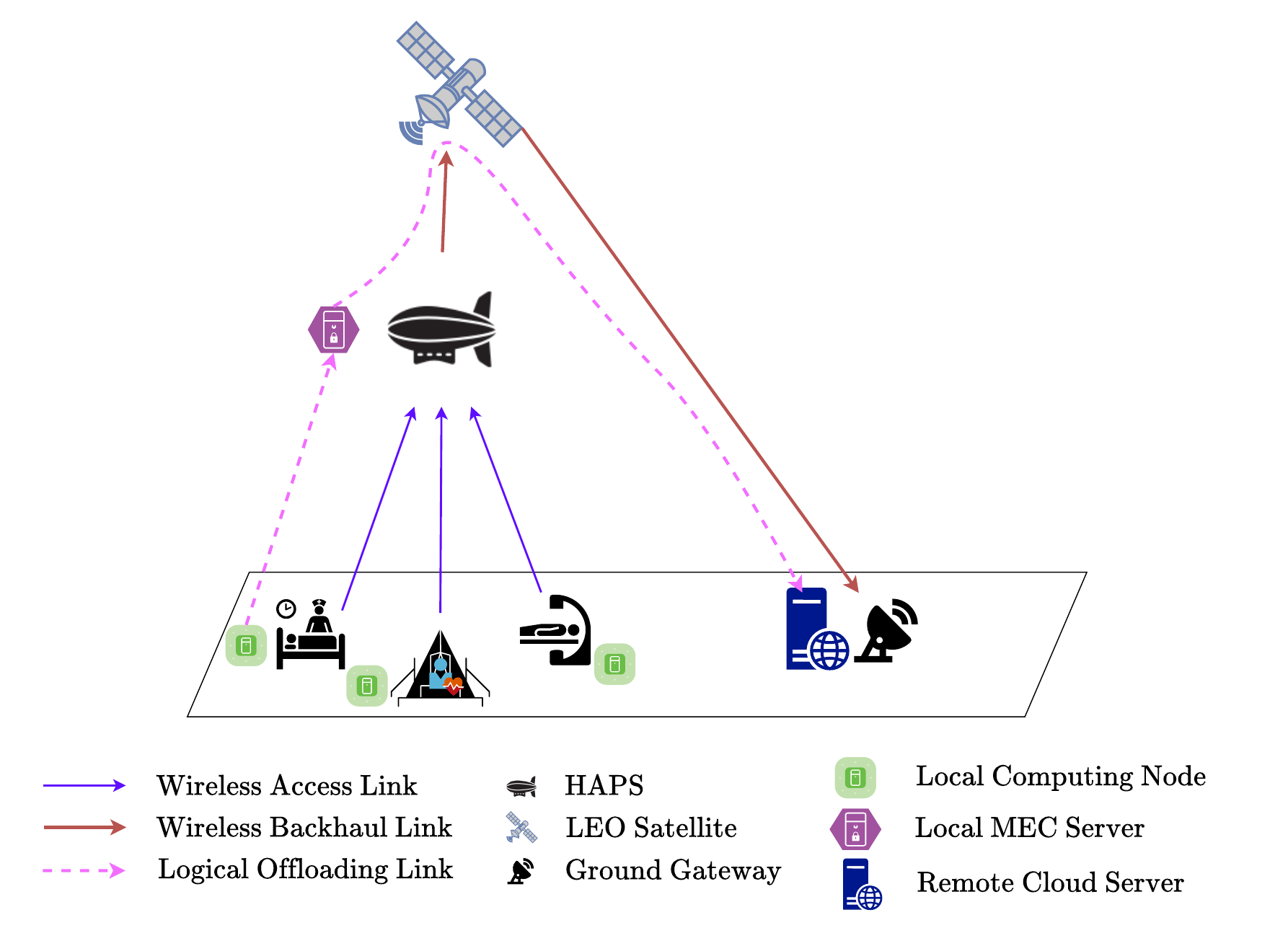}\vspace{-2em}
    \caption{System Model}
    \label{fig:sysmodel}
\end{figure}

Assume an ad-hoc healthcare deployment has been installed in a remote location, due to a general disaster situation. The healthcare ground devices (GDs), represented by the set $\mathcal{I}$ do not have access to a terrestrial network due to the geographical remoteness. Each GD gathers biosignal-related data from the patients, and, at a given snapshot of the system, generates a computing task $\pmb{\psi}_i = [d_i,\mu_i, \tau_i^{\maxt}]^T$ related to the data gathered, where $d_i$ is the number of bits of the task, $\mu_i$ is its computation density in CPU cycles per bit in Hz, and $\tau_i^{\maxt}$ is the maximum delay allowed for its completion. A local HAPS $h$ with MEC capabilities is deployed, which can compute tasks generated by the GDs, transmitted through a shared wireless channel of bandwidth $B_u$. The HAPS charges the GDs a price $c_{B_u}$ per unit bandwidth to access the wireless channel in the S band, and a price $c_i^{\mathrm{MEC}}$ to compute their corresponding tasks. Furthermore, the HAPS can connect to LEO satellite $s$ belonging to a different provider, through a wireless channel of bandwidth $B_h$ in the Ka band. The LEO satellite acts as a bridge to a remote MCC server, through a ground gateway $g$, which offers ample computing capabilities. We assume the HAPS does not have access to the terrestrial infrastructure given the remoteness of the location, thus, the satellite system charges the HAPS a price of $c_{B_h}$ per unit bandwidth to access the wireless channel, and a price $c_i^{\mathrm{MCC}}$ to compute the corresponding task $\pmb{\psi}_i$. Figure~\ref{fig:sysmodel} shows the system model considered. 
\vspace{-1em}

\subsection{Channel Model}

The HAPS has a uniform planar array (UPA) of $M_{h}^{d}$ antenna elements for GD access, and a UPA of $M_{h}^{u}$ antenna elements for access to the LEO satellite. The LEO satellite and the GW have UPAs of $M_{s}$ and $M_{g}$ antenna elements respectively. The IoMT-HAPS channels $\mathbf{h}_{i,h}$ 
% are in the 2.1 GHz LTE band, with an
includes an air-to-ground pathloss component given as in~\cite{art:Hourani_UAVChannels}, and a Rician small-scale fading component modeled as in~\cite{art:Cheng_RicianChans}. To receive the signals sent by the GDs, the HAPS employs a maximum-ratio combiner, having a channel gain of $g_{i,h} = ||\mathbf{h}_{i,h}||^2 $.

The links between the HAPS and the LEO satellite, and between the LEO satellite and the GW are orthogonal of bandwidth $B_h$. Given the spatial sparsity of the links, we assume LoS-MIMO channels, for which the corresponding channel gains are given as $g_{j,k} = M_jM_k \left(\frac{\lambda_k  G_{u,k}}{4\pi ||\mathbf{r}_j - \mathbf{r}_k||_2}\right) ^2$,
% \begin{align}
%     g_{j,k} = M_jM_k \left(\frac{\lambda_k  G_{u,k}}{4\pi ||\mathbf{r}_j - \mathbf{r}_k||_2}\right) ^2
% \end{align}
with $j\in\{h,s\}$, $k\in\{s,g\}$, $G_{u,k}$ includes the atmospheric losses due to weather phenomena, and $\lambda_k$ is the corresponding wavelength. Then, the communication rate between node $j\in\mathcal{I}\cup\{h,s\}$ and $k\in\{h,s,g\}$ is 
\begin{align}\label{eq:Riu}
    R_{j,k} =  \rho_jB_{j\rightarrow k} \log_2\left( 1 + \frac{g_{j,k}p_{j}}{\rho_jB_kN_0} \right),
\end{align}
where $\rho_j$ is the proportion of the bandwidth available used by node $j$, $B_{j\rightarrow k}\in\{B_u, B_h\}$ is the total bandwidth available for the link, $N_0$ is the noise power per unit bandwidth at reception, and $p_{j}$ is the corresponding transmit power. 

\subsection{Delay Model}

\subsubsection{Transmission Delay}
% \paragraph{Ground devices - HAPS}

The transmission delay experienced by offloading task $\pmb{\psi}_i$ from GD $i$ to the HAPS $h$ is given as $\tau_{i,h} = \frac{d_i}{ R_{i,h}}$.
% \begin{align}
%     \tau_{i,h} = \frac{d_i}{ R_{i,h}}.
% \end{align}
% \paragraph{UAV - HAPS}
% The transmission delay experienced by offloading the required tasks from UAV $u$ to the HAPS, is
% \begin{align}
%     \tau_{u,h} &= \frac{ \sum\limits_{i\in\mathcal{I}_u} \beta_{u,h}^i d_i}{R_{u,h}}.
% \end{align}
% where $\beta_{u,h}^i$ is a binary variable indicating if task $i$ is offloaded from UAV $u$ to the HAPS ($\beta_{u,h}^i=1$) or not ($\beta_{u,h}^i=0$). 
% \paragraph{HAPS/LEO - LEO/GW}
% From the HAPS or LEO satellite, to the LEO satellite or the GW, the round-trip propagation delay cannot be neglected, due to the large distances. 
In the delay experienced by offloading the corresponding tasks from $j\in\{h,s\}$ to $k\in\{s,g\}$ the round-trip propagation delay cannot be neglected, due to the large distances, and is given as $\tau_{j,k} = \frac{ 1}{R_{j,k}}\sum_{i\in\mathcal{I}_h} \beta_{h,s}^i d_i+ 2\tau_{j,k}^{\text{prop}}$, where $\tau_{j,k}^{\text{prop}} =\frac{1}{c}||\mathbf{r}_{j} - \mathbf{r}_{k}||_2$. 
% a sum of the transmission delay and the propagation delay, given as
% \begin{align}
%     \tau_{u,s} = \frac{ \sum\limits_{i\in\mathcal{T}_{u}^{s}} d_i}{R_{u,s}}.
% \end{align}
% \begin{align}
%     \tau_{j,k} &= \frac{ \sum\limits_{i\in\mathcal{I}_h} \beta_{h,s}^i d_i}{R_{j,k}}+ 2\underbrace{\frac{||\mathbf{r}_{j} - \mathbf{r}_{k}||_2}{c}}_{\tau_{j,k}^{\text{prop}}}.
% \end{align}
The offloading decision variable $\beta_{h,s}^i = 1$ if task $\pmb{\psi}_i$ is offloaded from $h$ to $s$ and $\beta_{h,s}^i = 0$ otherwise. Likewise, $\beta_{u,h}^i = 1$ if task $\pmb{\psi}_i$ is offloaded from GD $i$ to $s$ and $\beta_{u,h}^i = 0$ otherwise.
% \paragraph{LEO - GW}
% The tasks received at the LEO satellite are sent to a remote gateway within coverage of the LEO satellite for cloud processing. Likewise, the roundtrip propagation delay between the LEO satellite and the ground gateway is considered within the corresponding delay, given as
% % \begin{align}
% %     \tau_{u,s} = \frac{ \sum\limits_{i\in\mathcal{T}_{u}^{s}} d_i}{R_{u,s}}.
% % \end{align}
% \begin{align}
%     \tau_{s,g} &= \frac{ \sum\limits_{i\in\mathcal{I}_h} \beta_{h,s}^i d_i}{R_{s,g}}+ 2\underbrace{\frac{||\mathbf{r}_{g} - \mathbf{r}_{s}||_2}{c}}_{\tau_{s,g}^{\text{prop}}}.
% \end{align}

\subsubsection{Computing Delay}
Task $\pmb{\psi}_i$ can be computed locally at GD $i\in\mathcal{I}$, or remotely at the HAPS or the remote MCC. If node $k\in\mathcal{I}\cup\{h\} $ computes the task, the computing delay is $\tau_i^k = \frac{  d_{i} \mu_{i} }{f_i^k}$, 
% \begin{align}
%     \tau_i^k = \frac{  d_{i} \mu_{i} }{f_i^k}, 
% \end{align}
where $f_i^k$ are the resources allocated for the task by node $k$. If it is computed at the MCC, the processing delay is assumed as negligible, due to its vast computing resources. % at this server.

% Then, the delay of computing at each computing layer is given as
% \begin{align}
%     \tau_{\mathrm{Loc},i} &= \tau_{i}^{i}\\
%     \tau_{\mathrm{MEC},i} &= \tau_{i,u}+\tau_{u,h} + \tau_{i}^{h}\\
%     \tau_{\mathrm{MCC},i} &= \tau_{i,u}+\tau_{u,h}+\tau_{h,s}+\tau_{s,g}
% \end{align}

% Thus, the total delay per task is given as
% \begin{align}
%     \tau_i &= (1 - \beta_{u,h}^{i})\tau_{\mathrm{Loc},i} + \beta_{u,h}^{i}(1 - \beta_{h,s}^{i})\tau_{\mathrm{MEC},i} + \beta_{h,s}^{i}\tau_{\mathrm{MCC},i}
% \end{align}

\subsection{Cost Model}

The cost of computing or offloading a task is defined by the cost of bandwidth, the cost of remote computing and a delay cost to incentivize its prompt execution.

% In offloading, a cost due to the use of spectral resources is incurred. Furthermore, to incentivize the prompt computation of the tasks, we introduce a weighted delay cost.

\subsubsection{Cost for GDs}

% The GD has three costs associated to their generated task: the cost related to the delay of performing the task, the cost related to its computation by the local MEC server, and the cost related to the use of the spectrum. 
We suppose a virtual cost of $c_{\tau}$ per second spent in computing the task; that the ground devices pay $c_{B_u}$ per Hz of the use of the spectrum to the HAPS; and that the MEC server sets and advertises a cost $c_i^{\mathrm{MEC}}$ for computing the task. Thus, cost for GD $i$ is
% the cost incurred by the ground device related to the processing of its task $i$ is given as
% \begin{align}\nonumber
%     C_i^{\mathrm{loc}} &= (1-\beta_{u,h}^{i})c_{\tau}\tau_{i}^{i} \\
%     &+ \beta_{u,h}^{i}\left(c_{\tau}\tau_{i,h} + c_{B}\rho_iB_u + c_i^{\mathrm{MEC}} \right).
% \end{align}
\begin{align}\label{eq:Ciloc}\small
    C_i^{\mathrm{loc}} &= (1-\beta_{u,h}^{i})c_{\tau}\tau_{i}^{i} + \beta_{u,h}^{i}\left(c_{\tau}\tau_{i,h} + c_{B}\rho_iB_u + c_i^{\mathrm{MEC}} \right).
\end{align}
% where $\beta_{u,h}^{i} = 1$ if task $\pmb{\psi}_i$ is offloaded to the HAPS and $\beta_{u,h}^{i} = 0$ otherwise.

\subsubsection{Cost and Utility for MEC server}

The MEC server earns a profit from the offloading of a task from GD $i$. 
% In this context it incurs in a cost related to the delay of the task. 
If it offloads it to the MCC through the LEO satellite, it incurs a processing cost and bandwidth cost. The LEO charges the HAPS $c_{B_u}$ per Hz of the use of the spectrum. Thus, the total cost incurred by the HAPS is
% \begin{align}\nonumber
%     C^{\mathrm{MEC}} &= c_{B_h}\rho_{B}B_h - \sum_{i\in\mathcal{I}}\beta_{u,h}^{i}\left(c_{B}\rho_iB_u + c_i^{\mathrm{MEC}}\right) \\
%     \nonumber & + \sum_{i\in\mathcal{I}}(\beta_{u,h}^{i} - \beta_{h,s}^{i})c_{\tau}\tau_{i}^{h} \\
%     & + \sum_{i\in\mathcal{I}}\beta_{h,s}^{i} \left( c_{\tau}\left(\tau_{h,s} + \tau_{s,g}\right)+ c_i^{\mathrm{MCC}}  \right)
%     \label{eq:MEC_cost}
% \end{align}
\begin{align}\label{eq:MEC_cost}\small
    &C^{\mathrm{MEC}} = c_{B_h}\rho_{B}B_h - \sum_{i\in\mathcal{I}}\beta_{u,h}^{i}\left(c_{B}\rho_iB_u + c_i^{\mathrm{MEC}}\right) \\
    \nonumber  & + \sum_{i\in\mathcal{I}}(\beta_{u,h}^{i} - \beta_{h,s}^{i})c_{\tau}\tau_{i}^{h}  + \sum_{i\in\mathcal{I}}\beta_{h,s}^{i} \left( c_{\tau}\left(\tau_{h,s} + \tau_{s,g}\right)+ c_i^{\mathrm{MCC}}  \right).
\end{align}

% \begin{align}
%     C_i^{\mathrm{MEC}} &= - \beta_{u,h}^{i}\left(c_{B}\rho_iB_u + c_i^{\mathrm{MEC}}\right) \\
%     & + (\beta_{u,h}^{i} - \beta_{h,s}^{i})c_{\tau}\tau_{i}^{h} \\
%     & + \beta_{h,s}^{i} \left( c_{\tau}\left(\tau_{h,s} + \tau_{s,g}\right)+ c_i^{\mathrm{MCC}} + c_{B_h}\rho_{B}B_h \right)
% \end{align}

\subsubsection{Utility for Cloud Server}

% The cloud server is considered to have a vast amount of computing resources, for which 
% The computing time of its offloaded tasks is, thus, 
The cloud server benefits from the computing of its tasks as
\begin{align}\label{eq:MCC}\small
    C^{\mathrm{MCC}} &= - c_{B_h}\rho_{B}B_h - \sum_{i\in\mathcal{I}}\beta_{h,s}^{i}c_i^{\mathrm{MCC}} 
\end{align}

\section{Problem Formulation}
% \section{Optimization Problem}

Each entity in the system seeks to maximize its utility. The satellite system computes the optimal cost per task to the HAPS, 
% such that the HAPS is incentivized to offload it. This is 
represented by the following optimization problem:
\begin{subequations}\label{eq:opt_MCC}
{
\small
\begin{alignat}{3}\label{eq:MCC_1}
\mathcal{P}^{\mathrm{MCC}}:\;\;\; &\min_{\{c_i^{\mathrm{MCC}}\}} & &  C^{\mathrm{MCC}}  \\
% \label{eq:MCC_2}        &   &\Cb{1}^{c}:\quad&  \tau_{i,h} + \tau_{h,s} + \tau_{s,g} \leq \tau_{i}^{\maxt}  \;\; \forall i\in\mathcal{I} \\
% \label{eq:MCC_3}        &   &\Cb{2}^{c}:\quad&  c_{\tau}\tau_{i}^{h} \geq c_{\tau}\left(\tau_{h,s} + \tau_{s,g}\right)+ c_i^{\mathrm{MCC}}  \;\; \forall i\in\mathcal{I} \\
\label{eq:MCC_4}        &\mathrm{s.t.}   &\Cb{1}^{c}:\quad& c_i^{\mathrm{MCC}} \geq 0  \;\;
\end{alignat}
}\end{subequations}
% where $\Cb{1}^{c}$ enforces the maximum delay constraint for the task, $\Cb{2}^{c}$ incentivizes the offloading of the task from the HAPS to the MCC, and $\Cb{3}^{c}$ specifies that costs are non-negative.

% $ + \sum_{i\in\mathcal{I}}(\beta_{u,h}^{i} - \beta_{h,s}^{i})c_{\tau}\tau_{i}^{h}  + \sum_{i\in\mathcal{I}}\beta_{h,s}^{i} \left( c_{\tau}\left(\tau_{h,s} + \tau_{s,g}\right)+ c_i^{\mathrm{MCC}}  \right).$

The HAPS computes the optimal offloading strategy to the cloud and the corresponding share of the bandwidth, as well as the optimal cost per task to each GD that solves the following optimization problem
\begin{subequations}\label{eq:opt_MEC}
{
\small
\begin{alignat}{3}\label{eq:MEC_1}
\mathcal{P}^{\mathrm{MEC}}:\;\;\; &\min_{\substack{\{c_i^{\mathrm{MEC}}\}\\\{\beta_{u,s}^{i}\},\rho_B}} & & C^{\mathrm{MEC}}  \\ 
\label{eq:MEC_2}        &   &\Cb{1}^{m}:\quad&  \tau_{i,h} + \left(1-\beta_{u,s}^{i}\right)\tau_{i}^{h}  \\
\nonumber        &   &\quad&  +\beta_{u,s}^{i}\left( \tau_{h,s} + \tau_{s,g}\right) \leq \tau_{i}^{\maxt}  \;\;  \forall i\in\mathcal{I}  \\
% \label{eq:MEC_In}       &   &\Cb{2}^{m}:\quad&  c_{\tau}\tau_{i}^{i} \geq c_{\tau}\tau_{i,h} + c_{B}\rho_iB_u + c_i^{\mathrm{MEC}}  \;\;  \forall i\in\mathcal{I}  \\
\label{eq:MEC_3}        &   &\Cb{2}^{m}:\quad&  \beta_{h,s}^{i} \in \{0,1\},  \;\;  \forall i\in\mathcal{I}  \\
\label{eq:MEC_4}        &   &\Cb{3}^{m}:\quad&  0\leq\rho_{B}\leq 1.  \;\; \\
\label{eq:MEC_5}        &   &\Cb{4}^{m}:\quad&  c_i^{\mathrm{MEC}}\geq 0.  \;\; \forall i\in\mathcal{I}
\end{alignat}
}\end{subequations}
where $\Cb{1}^m$ enforces the maximum delay constraint, $\Cb{2}^m$ 
% incentivizes the offloading of tasks from the GDs, $\Cb{3}^m$ 
indicates that the offloading decisions are binary, $\Cb{3}^m$ indicates that a non-negative proportion of the bandwidth to the MCC is used, and $\Cb{4}^m$ indicates that costs are non-negative.
% $(1-\beta_{u,h}^{i})c_{\tau}\tau_{i}^{i} + \beta_{u,h}^{i}\left(c_{\tau}\tau_{i,h} + c_{B}\rho_iB_u + c_i^{\mathrm{MEC}} \right)$

Each GD $i\in\mathcal{I}$ maximizes its own utility, by determining their offloading strategy and portion of the bandwidth, that solves the following optimization problem
\begin{subequations}\label{eq:opt_loc}
{
    \small
    \begin{alignat}{3}\label{eq:loc_1}
    \mathcal{P}_i^{\mathrm{loc}}:\;\;\; &\min_{\beta_{u,h}^{i},\rho_i} & &C_i^{\mathrm{loc}}  \\ 
    \label{eq:loc_2}        &   &\Cb{1}^{l}:\quad&  \left(1 - \beta_{u,h}^{i}\right)\tau_{i}^{i} + \beta_{u,h}^{i}\left(\tau_{i,h} + \tau_{i}^{h} \right) \leq \tau_{i}^{\maxt}  \;\;    \\
    \label{eq:loc_3}        &   &\Cb{2}^{l}:\quad&  \beta_{u,h}^{i} \in \{0,1\},  \;\;  \\
    \label{eq:loc_4}        &   &\Cb{3}^{l}:\quad&  \rho_{i} + \sum_{\substack{j\in\mathcal{I} \\ j\neq i}}\rho_{j} \leq 1 \\
    \label{eq:loc_5}        &   &\Cb{4}^{l}:\quad&  0\leq\rho_{i} \leq 1 ,  \;\; 
    \end{alignat}
}
\end{subequations}
where $\Cb{1}^{l}$ indicates the maximum delay constraint of the task, and considers only offloading to the HAPS, without information of the space segment. $\Cb{2}^{l}$ indicates that the offloading is binary, $\Cb{3}^{l}$ limits the total bandwidth of the channel to the HAPS across GDs, and $\Cb{4}^{l}$ indicates that a non-negative proportion of the bandwidth to the HAPS is used.

\section{Proposed Solution}

% As a contribution of our work, we present different sets of solutions for different model metrics. The spectrum allocation problem, as a more regulated resource, is going to be handled in subsection \ref{sec:Spectrum} considering a central coordination point for policy setting. As for cost and offloading policies, according to the more dynamic system setting and the availability of multiple options for service provision, we propose a different point of view. 

% Subsections \ref{sec:priceoofload} and \ref{sec:Spectrum} will look into the problem of policy setting for offloading-based pricing spectrum allocation respectively.

In \ref{sec:priceoofload} we propose selfish price-setting policies for offloading, whereas in \ref{sec:Spectrum} we derive the optimal spectrum allocation.

\subsection{Price Setting and Offloading}\label{sec:priceoofload}

Solving $\mathcal{P}^{\mathrm{MCC}}$ and $\mathcal{P}^{\mathrm{MEC}}$ in isolation will cause the prices to go to infinity, minimizing their cost. However, the inherent competition over utility prevents the parties from doing so. To capture this, we model this scenario as a competition between tiers,
% This models a real application scenario where each processing node seeks to maximize its utility 
% % , considering their local cost and trying to 
% by optimizing their policies accordingly. 
 and develop a game-theoretic framework to solve it, where the cost and offloading problems are solved simultaneously. 
The players are the three tiers with computation capabilities. To separate the interactions, we decompose \eqref{eq:MEC_cost}. The corresponding partial cost for the interaction between GD and MEC is given as $C^{\mathrm{MEC,loc}} = \sum_{i\in\mathcal{I}}\beta_{u,h}^{i}\left(c_{\tau}\tau_{i}^{h}-c_{B}\rho_iB_u - c_i^{\mathrm{MEC}}\right)$. 
% \begin{align}
%     C^{\mathrm{MEC,loc}} &=
%     \begin{aligned}[t]
%         & - \sum_{i\in\mathcal{I}}\beta_{u,h}^{i}\left(c_{B}\rho_iB_u + c_i^{\mathrm{MEC}}\right) \\
%         & + \sum_{i\in\mathcal{I}}\beta_{u,h}^{i} c_{\tau}\tau_{i}^{h}
%     \end{aligned}
%     \label{eq:MEC_LOC_cost}        
% \end{align}
% \begin{align}\label{eq:MEC_LOC_cost}
%     C^{\mathrm{MEC,loc}} &= - \sum_{i\in\mathcal{I}}\beta_{u,h}^{i}\left(c_{B}\rho_iB_u + c_i^{\mathrm{MEC}}\right) + \sum_{i\in\mathcal{I}}\beta_{u,h}^{i} c_{\tau}\tau_{i}^{h}    
% \end{align}
% \begin{align}\label{eq:MEC_LOC_cost}
%     C^{\mathrm{MEC,loc}} &= \sum_{i\in\mathcal{I}}\beta_{u,h}^{i}\left(c_{\tau}\tau_{i}^{h}-c_{B}\rho_iB_u - c_i^{\mathrm{MEC}}\right) 
% \end{align}
Likewise, the partial cost corresponding to the interaction between the HAPS and the MCC is given as $C^{\mathrm{MEC,MCC}} = c_{B_h}\rho_{B}B_h  + \sum_{i\in\mathcal{I}} \beta_{h,s}^{i} 
            \left( c_{\tau}\left(\tau_{h,s} + \tau_{s,g} - \tau_{i}^{h} \right) + c_i^{\mathrm{MCC}} \right)$.
% The next equation will model the HAPS cost interacting with the cloud server, located in the ground gateway which is accessed through LEO satellite. To achieve this, we decompose \eqref{eq:MEC_cost} as 
% \begin{align}
%     C^{\mathrm{MEC,MCC}} &=
%     \begin{aligned}[t]
%         & c_{B_h}\rho_{B}B_h \\
%         & + \sum_{i\in\mathcal{I}}\left( - \beta_{h,s}^{i} c_{\tau} \tau_{i}^{h} \right) \\
%         & + \sum_{i\in\mathcal{I}}\left( \beta_{h,s}^{i} 
%             \left( c_{\tau}\left(\tau_{h,s} + \tau_{s,g}\right) + c_i^{\mathrm{MCC}} \right)
%           \right)
%     \end{aligned}
%     \label{eq:MEC_MCC_cost}    
% \end{align}
% \begin{align}
%     C^{\mathrm{MEC,MCC}} &=
%     \begin{aligned}[t]
%         & c_{B_h}\rho_{B}B_h \\
%         & + \sum_{i\in\mathcal{I}} \beta_{h,s}^{i} 
%             \left( c_{\tau}\left(\tau_{h,s} + \tau_{s,g}\right) + c_i^{\mathrm{MCC}} - c_{\tau} \tau_{i}^{h} \right)
%     \end{aligned}
%     \label{eq:MEC_MCC_cost}    
% \end{align}
% \begin{align}\label{eq:MEC_MCC_cost}    
%     C^{\mathrm{MEC,MCC}} &= c_{B_h}\rho_{B}B_h \\
%     \nonumber    & + \sum_{i\in\mathcal{I}} \beta_{h,s}^{i} 
%             \left( c_{\tau}\left(\tau_{h,s} + \tau_{s,g} - \tau_{i}^{h} \right) + c_i^{\mathrm{MCC}} \right)
% \end{align}
The interactions are analyzed in an extensive-form game. 

Considering~\eqref{eq:Ciloc}, ~\eqref{eq:MCC}, $C^{\mathrm{MEC,loc}}$ and $C^{\mathrm{MEC,MCC}}$, it can be seen that GD has control over offloading and the MCC over pricing policy, whereas the HAPS can decide over the two policies: offloading to the MCC, pricing to the GDs. 
% The policies are set considering the delay constraints, crucial for medical applications.
% In this hierarchical architecture, the MCC sets the pricing policy and communicates it to the HAPS, which complies with the pricing.
As the first round of policy setting, the MCC sets its strategy knowing the response from HAPS, which is modeled as:
\begin{equation}\label{Game}  
\begin{aligned}
\min_{\{c_i^{\text{MCC}}\}} \quad 
& C^{\text{MCC}}
\end{aligned}
\end{equation}

% After reporting the price, the HAPS sets the variables related to the game-theoretic interaction with ground gateway.

% The interactions are analyzed in an extensive-form game. In this hierarchical architecture, the MCC sets the corresponding prices with knowledge of the offloading response from the HAPS to incentivize offloading as
% \begin{align}
%     c_i^{\text{MCC}} &= \left[c_{\tau} \left( \tau_{i}^{h} - \tau_{h,s} - \tau_{s,g} \right)\right]^+
% \end{align}

The MCC reports the chosen price to the HAPS, which then sets the corresponding offloading variables $\beta_{h,s}^i$ with fixed bandwidth allocation as
\begin{align}\label{prob:min_beta_hs}
\min_{\{\beta_{h,s}^i\}} &\quad 
C^{\mathrm{MEC,MCC}} \left(c_i^{\mathrm{MCC}}\right)\\
\nonumber \mathrm{s.t.}&\quad  \beta_{h,s}^{i} \in \{0,1\},  \;\;  \forall i\in\mathcal{I}  
\end{align}
The feasibility constraints will be enforced on the solution of the game theoretic framework. 
% This notation states that the optimization problem is solved considering availability of the knowledge from the previous stage of policy setting, here meaning the availability of the price of ground gateway.
Using backward induction, as the MCC knows the following action from the HAPS, it will set its policies accordingly.

The interaction between HAPS and the GDs is modelled by the following problem
\begin{equation}
    \begin{aligned}
        \min_{\{c_i^{\mathrm{MEC}}\}} \quad 
        C^{\mathrm{MEC,loc}}
        &
    \end{aligned}
    \label{game2}
\end{equation}
Following the resulting policy, the GDs determine the offloading strategy as
% wether to offload or not, deciding over the optimization problem:
\begin{align}\label{prob:min_beta_local_cost}
        \min_{\{\beta_{u,h}^i\}} &\quad 
        C_i^{\mathrm{loc}} \left(c_i^{\mathrm{MEC}}\right)\\
 \nonumber\mathrm{s.t.}&\quad  \beta_{u,h}^{i} \in \{0,1\},  \;\;  \forall i\in\mathcal{I}    
\end{align}

As the equations are affine with respect to the values of $\beta$ at different tiers, the optimization will yield a marginal value. As for the price policies, players will try to minimize their cost, setting the boundary values 
% with also ensures seamless service, 
where offloading is not interrupted. 
A further bound is derived by considering the maximum delay constraint.
% Adding the delay constraints required for IoMT applications, another operational bound is derived. 
Considering the game theoretic solution of \eqref{Game} \eqref{prob:min_beta_hs} we have
\begin{align}\label{MCC1}
    \hat{c}_i^{\text{MCC}} &< c_{\tau}\left[  \tau_{i}^{h} - \tau_{h,s} - \tau_{s,g} \right]^+
\end{align}

Adding the delay constraint \eqref{eq:MEC_2} we have
\begin{align}\label{MCC2}
   \Bar{c}_i^{\text{MCC}}  \le c_{\tau}\left(\tau_i^{\max} + \tau_i^{h} - 2\,\tau_{h,s} - 2\,\tau_{s,g} - \tau_{i,h} \right)
\end{align}
Thus, the price policy is given as
% that also satisfies the constraints is
\begin{align}
\label{C_MCC_complete}
c_i^{\text{MCC}} = \min\{\hat{c}_i^{\text{MCC}},\Bar{c}_i^{\text{MCC}}\}
\end{align}

For the second game, a similar approach is followed, considering equations\eqref{game2},  \eqref{prob:min_beta_local_cost} and \eqref{eq:loc_2}
\begin{align}\label{MEC1}
    \hat{c}_i^{\text{MEC}} &< \left[c_{\tau}  \tau_{i}^{i} - c_{\tau}  \tau_{i,h} - c_{B} \rho_{i} B_{u} \right]^+
\end{align}
\begin{align}\label{MEC2}
 \Bar{c}_i^{\text{MEC}}  \le c_{\tau}\left( \tau_{i}^{i} - \tau_{i}^{h} - 2\,\tau_{i,h} + \tau_i^{\max}\right) - c_{B} \rho_{i} B_{u}
\end{align}
Forming the final policy as
\begin{align}
\label{C_MEC_complete}
c_i^{\text{MEC}}  = \min\{\hat{c}_i^{\text{MEC}},\Bar{c}_i^{\text{MEC}}\}
\end{align}

It is worth noting that the delay constraint ensures service provision, even at the cost of higher expenses for the players.

\subsection{Spectrum Allocation}\label{sec:Spectrum}

\subsubsection{HAPS-GW}

Let $\mathcal{I}^{\mathrm{s}}$ be the set of tasks offloaded from the HAPS to the MCC. We assume a constant transmit power per unit bandwidth from the HAPS and the LEO satellite $\hat{p}_{i,k} = \frac{p_{i,k}}{\rho_BB_h}$. Then, after some algebraic manipulation we can write \eqref{eq:MEC_cost} as a function of $\rho_{B}$ as $C_i^{\mathrm{MEC}} = \frac{a_h}{\rho_B} + b_h \rho_{B}$
% \begin{align}
%     C_i^{\mathrm{MEC}} &= \frac{a_h}{\rho_B} + b_h \rho_{B}
% \end{align}
where
{\small
\begin{align}
    a_h &= \frac{c_{\tau}I^{\mathrm{s}}\sum\limits_{i\in\mathcal{I}^{\mathrm{s}}}  d_i}{B_h}\left(\frac{1}{ \log_2\left( 1 + \frac{g_{h,s}\hat{p}_{h,s}}{N_0} \right)} + \frac{ 1}{\log_2\left( 1 + \frac{g_{s,g}\hat{p}_{s,g}}{N_0} \right)} \right)\\
    b_h &= c_{B_h}B_h
\end{align}}
which is a convex function over $\rho_{B}>0$. By taking its first derivative and equaling it to zero, its optimal point is given at $\rho_{B}^{\text{opt}} = \sqrt{\frac{a_h}{b_h}}$. Furthermore, to ensure that constraint $\Cb{1}^{m}$ is met, it must hold that 
\begin{align}
    \rho_{B} &\geq \underbrace{\frac{a_h}{c_{\tau}I^s\min\limits_{i\in\mathcal{I}^s}\left\{\tau_{i}^{\maxt} - \left(\tau_{i,h}+2\left(\tau_{h,s}^{\text{prop}} + \tau_{s,g}^{\text{prop}}\right)\right)\right\}}}_{\rho_{B,1}^{\mint}}
\end{align}

To ensure that the cost for offloading does not go above the cost of local execution, it needs to hold that
\begin{align}
    \frac{a_h}{\rho_B} + b_h\rho_B &\leq \underbrace{\sum_{i\in\mathcal{I}^s}\left(c_{\tau}\tau_{i}^{h} - c_i^{\mathrm{MCC}}\right)}_{\sigma}
\end{align}
This expression can be converted into the quadratic convex form $b_h\rho_B^2 - \sigma \rho_B + a_h \leq 0$. By finding the cuts to zero of the function, we can convert this into the following equivalent linear constraint
\begin{align}
    \underbrace{\frac{\sigma - \sqrt{\sigma^2 -4b_ha_h}}{2b_h}}_{\rho_{B,2}^{\mint}} \leq &\rho_B \leq \underbrace{\frac{\sigma + \sqrt{\sigma^2 -4b_ha_h}}{2b_h}}_{\rho_{B}^{\maxt}}
\end{align}
which implies that $\sigma \geq 2\sqrt{b_ha_h}$ must hold. We set $\rho_{B}^{\mint} = \max{\{\rho_{B,1}^{\mint},\rho_{B,2}^{\mint}\}}$. Then, the bandwidth allocation algorithm for the HAPS-LEO-GW is given as in Algorithm~\ref{alg:rhoB}.

\SetKwBlock{RepeatE}{repeat}{}
% \begin{figure}
\begin{algorithm}
{\small
    \caption{ HAPS-LEO-GW Bandwidth allocation }\label{alg:rhoB}
    \Repeat{$\sigma \geq 2\sqrt{b_ha_h}$}{
        Set $\beta_{h,s}^{i^{*}}=0$ for $i^{*} = \argmin\left\{c_{\tau}\tau_{i}^{h} - c_i^{\mathrm{MCC}}\right\}$.\;
    }
    \Repeat{$ \rho_{B}^{\mint} \leq 1 $ \textbf{and} $\rho_{B}^{\mint}\leq \rho_{B}^{\maxt}$ }{
        Set $\beta_{h,s}^{i^{*}}=0$ for $i^{*} = \argmin\left\{\tau_{i}^{\maxt} - \tau_{i,h}\right\}$.\;
    }
    \uIf{$ \rho_B^{\mint}\leq \rho_B^{*} \leq \rho_B^{\maxt}$}{
        Set $ \rho_B = \rho_B^{*} $. \;
      }
      \Else{
        Set $\rho_B = \argmin\left\{C_i^{\mathrm{MEC}}\left(\rho_B^{\mint}\right), C_i^{\mathrm{MEC}}\left(\rho_B^{\maxt}\right)\right\}$ \;
      }
    }
\end{algorithm}

% For the offloading to preserve feasibility for all tasks, it must hold that $\rho_{B}^{\mint}\leq 1$. If $\rho_{B}^{\mint}> 1$, then task $i^* = \argmin_{i\in\mathcal{I}^s}\left\{\tau_{i}^{\maxt} - \left(\tau_{i,h}+2\left(\tau_{h,s}^{\text{prop}} + \tau_{s,g}^{\text{prop}}\right)\right)\right\}$ is dropped from $\mathcal{I}^s$ consecutively, until $\rho_{B}^{\mint}\leq 1$ is met.
% % We assume this is enforced by the offloading solution. 
% Then, the optimal HAPS-LEO-GW spectrum allocation is given as
% \begin{align}
%     \rho_{B}^{*} = \min\left\{\max\left\{\sqrt{\frac{a_h}{b_h}},\rho_{B}^{\mint}\right\},1\right\}
% \end{align}
% % \begin{align}
% %     \rho_{B}^{*} = \min\left\{\max\left\{\sqrt{\frac{a_h}{b_h}},\frac{a_h}{c_{\tau}I^s\min\limits_{i\in\mathcal{I}^s}\left\{\tau_{i}^{\maxt} - \left(\tau_{i,h}+2\left(\tau_{h,s}^{\text{prop}} + \tau_{s,g}^{\text{prop}}\right)\right)\right\}}\right\},1\right\}
% % \end{align}

\subsubsection{GD-HAPS}

Let $\mathcal{I}^{h}$ be the set of tasks offloaded from the GDs to the HAPS. Similar to the HAPS-GW link, the optimal solution of the objective function for GD $i\in\mathcal{I}^h$ is given as $\rho_{i}^{*} = \sqrt{\frac{a_i}{b_i}}$, where 
\begin{align}
    a_i &= \frac{c_{\tau}d_i}{ B_u \log_2\left( 1 + \frac{g_{i,h}\hat{p}_{i,k}}{N_0} \right)}\\
    b_i &= c_{B}B_u
\end{align}
This solution is generally not jointly feasible for all GDs, as $\Cb{2}^{l}$ is not generally met. If the total amount of spectral resources are fully occupied, and the local optimal for a given GD is not met, requesting more spectral resources would increase the cost of the other GDs. This is a Pareto-optimal point. Thus, we find the solution as a fair Pareto-optimal point with the following min-max optimization problem:
\begin{subequations}\label{eq:opt_loc_all}
{
\small
\begin{alignat}{3}\label{eq:loc_all_1}
\mathcal{P}^{\mathrm{loc}}_{R}:\;\;\; &\min_{\{\rho_i\}} & & \max_{i\in\mathcal{I}^{h}} C_i^{\mathrm{loc}}  \\ 
\label{eq:loc_all_2}        &   &\Cb{1}^{L}:\quad&  \rho_{i} \geq \frac{a_i}{c_{\tau}\left(\tau_{i}^{\maxt}-\tau_{i}^{h}\right)} = \rho_{i,1}^{\mint}  \\
\label{eq:loc_all_3}        &   &\Cb{2}^{L}:\quad&  \sum_{i\in\mathcal{I}^h}\rho_{i} \leq 1  \\
\label{eq:loc_all_4}        &   &\Cb{3}^{L}:\quad&  0\leq\rho_{i} \leq 1 \qquad \forall i\in\mathcal{I}^{h}  \;\; \\
\label{eq:loc_all_5}        &   &\Cb{4}^{L}:\quad&  c_{\tau}\tau_{i,h} + c_{B}\rho_iB_u + c_i^{\mathrm{MEC}} \leq c_{\tau}\tau_{i}^{i} \qquad \forall i\in\mathcal{I}^{h}  \;\; 
\end{alignat}
}\end{subequations}
Constraint $\Cb{4}^{L}$ is needed so that tasks offloaded are only the ones that benefit from non-local computing, and can be written as $\frac{a_i}{\rho_i} + b_i\rho_i  \leq c_{\tau}\tau_{i}^{i} - c_i^{\mathrm{MEC}} $. The problem can be written in epigraph form as \begin{subequations}\label{eq:opt_loc_all2}
{
\small
\begin{alignat}{3}\label{eq:loc_all2_1}
\mathcal{P}^{\mathrm{loc}}:\;\;\; &\min_{\{\rho_i\},\eta} & & \eta  \\ 
\label{eq:loc_all2_2}        &   &\Cb{1}^{LE}:\quad& \frac{a_i}{\rho_i} + b_i\rho_i  \leq \eta \qquad \forall i\in\mathcal{I}^{h} \\
\label{eq:loc_all2_3}        &   &\quad&  \Cb{1}^{L},\Cb{2}^{L},\Cb{3}^{L},\Cb{4}^{L} 
\end{alignat}
}\end{subequations}
Constraint $\Cb{4}^{L} $ can be disregarded by ensuring $\eta \leq \min\limits_{i\in\mathcal{I}^h}\left\{ c_{\tau}\tau_{i}^{i} - c_i^{\mathrm{MEC}} \right\} = \eta^{\maxt}$. Furthermore, similar to the previous section, constraint $\Cb{1}^{LE}$ can be written as 
\begin{align}
    \underbrace{\frac{\eta - \sqrt{\eta^2 -4b_ia_i}}{2b_i}}_{\rho_{i,2}^{\mint}} \leq &\rho_i \leq \underbrace{\frac{\eta + \sqrt{\eta^2 -4b_ia_i}}{2b_i}}_{\rho_{i}^{\maxt}}
\end{align}
which implies the constraint $\eta \geq 2 \max\limits_{i\in\mathcal{I}^h} \left\{\sqrt{b_ia_i}\right\} = \eta^{\mint}$. We set $\rho_{i}^{\mint} = \max\{\rho_{i,1}^{\mint},\rho_{i,2}^{\mint}\}$. With this in mind, we can define a bisection-based algorithm to find the appropriate $\eta$ value values, which runs for a set number of iterations $N_{\nu}$, and guarantees convergence within a precision of $\varepsilon = 2^{-N_{\nu}}\left(\eta^{\maxt} - \eta^{\mint}\right)$. From this the optimal $\rho_i$ values can be found. The GD-HAPS bandwidth allocation algorithm is presented in Algorithm~\ref{alg:rhoi}.
\begin{algorithm}
{\small
    \caption{ GD-HAPS Bandwidth allocation }\label{alg:rhoi}
    \Repeat{$\sum_{i\in\mathcal{I}^{h}}\rho_{i,1}^{\mint} \leq 1$}{
        Set $\beta_{u,h}^{i^{*}}=0$ for $i^{*} = \argmin\left\{\tau_{i}^{\maxt} - \tau_{i}^{h}\right\}$.\;
    }
    \Repeat{$\sum_{i\in\mathcal{I}^{h}}\rho_{i,2}^{\mint}\left(\eta^{\maxt}\right) \leq 1$ \textbf{and} $\eta^{\mint}\leq \eta^{\maxt}$ }{
        Set $\beta_{u,h}^{i^{*}}=0$ for $i^{*} = \argmin\left\{c_{\tau}\tau_{i}^{h} - c_i^{\mathrm{MEC}}\right\}$.\;
    }
    Find $\eta^{*} $ through bisection between $\eta^{\mint}\leq\eta\leq \eta^{\maxt}$ such that
    \begin{align}\nonumber
        \sum_{i\in\mathcal{I}^{h}}\max\left\{\frac{\eta - \sqrt{\eta^2 -4b_ia_i}}{2b_i} , \frac{a_i}{c_{\tau}\left(\tau_{i}^{\maxt}-\tau_{i}^{h}\right)}\right\} \leq 1
    \end{align}\;
    For all $i\in\mathcal{I}^h$
    \uIf{$ \rho_i^{\mint}\leq \rho_i^{*} \leq \rho_i^{\maxt}$}{
        Set $ \rho_i = \rho_i^{*} $. \;
      }
      \Else{
        Set $\rho_i = \argmin\left\{C_i^{\mathrm{loc}}\left(\rho_i^{\mint}\right), C_i^{\mathrm{loc}}\left(\rho_i^{\maxt}\right)\right\}$ 
      }
    }
\end{algorithm}

\section{Results}\label{sec:Results}

% We analyze the costs obtained by each tier after running the proposed algorithms.  

\subsection{Communication Parameters}

The carrier frequency for the HAPS-LEO link is in the Ka band, chosen as $28$GHz, with bandwidth $B_h=100$MHz~\cite{rep:3gpp_38.101-5}. For the IoT-HAPS access, we assume LTE-M enabled IoMT devices working in LTE channel~\cite{rep:3gpp_21.914}, we choose a subchannel bandwidth of $B_u = 1.4$MHz, accommodating $|\mathcal{I}|=14$ GDs per snapshot, over the carrier at $2.1$GHz. The nodes are located around the center of coordinates, while the HAPS, LEO satellite and gateway are located at $\mathbf{r}_h = [0,0,20]$km, $\mathbf{r}_s = [0,5,500]$km, and $\mathbf{r}_g = [0,10,0]$km respectively. We consider the MEC at the HAPS has $F_h = 10$GHz of computing resources, and transmit powers of $p_i = 500$mW, $p_h = 2$W and $p_s = 1$W, over their total correspondent bandwidth of $B_u = 20$MHz and $B_h = 200$MHz. For comparison, we set the bandwidth cost as $c_{B_u}= c_{B_h} = 1\times 10^{-13}$.

% We assume an LTE-M compliant communication rate of $1$Mbps from the IoMT devices to the HAPS, and 

\subsection{Healthcare Task Model}
We consider three types of healthcare-related data
\begin{itemize}
    \item Large payload medical data, such as echocardiogram data, as 640x480 pixels images, with 8 bits per pixel, acquired at 1Hz, and utilizing 3:1 lossless compression~\cite{web:echocard}, resulting in a generation of data at the rate of 819.2 Kbps.
    \item Medium payload medical data, such as electrocardiogram (ECG) data generated at 500 Hz from 12 leads, represented in 32 bits each~\cite{data:ECG}, generating data at 192 Kbps rate.
    \item Low payload medical data, such as photoplethysmography (PPG) data from wearable devices, generated at 64 Hz in single-precision doubles~\cite{data:PPG}, generating data at 2.048 Kbps. 
\end{itemize}

For comparison, we take uniform time intervals for data offloading of 100 ms over a 1 Mbps data rate, which results in packets of approximately 82 Kbs, 20 Kbs and 200 bits respectively. Moreover, to account for variations in the data gathering of the devices and decisions to offload certain samples or not, we consider that the size of the tasks follow normal distributions around those values. Furthermore, we consider a high computing density of $\mu_i = 500$ cycles per bit for the large payload data, and a small computing density of $\mu_i=50$ cycles per bit for the medium and low payload data. Moreover, we choose maximum delays constraints of 
% $\tau_i^{\maxt} = 30$s, $\tau_i^{\maxt} = 5$s and $\tau_i^{\maxt} = 0.1$s
$\tau_i^{\maxt} \in\{500,50,1\}$ ms
for the large, medium, and low payload data, respectively. 

\subsection{Performance Evaluation}
 
% % (https://physionet.org/content/kurias-ecg/1.0/)

% photoplethysmography (PPG) data from a wearable device, generated at 64 Hz in single-precision doubles, generating data at 2048 bps
% % (https://physionet.org/content/big-ideas-glycemic-wearable/1.1.2/)

% Echocardiogram data, generated in the form of images comprised of 640x480 pixels, each represented by 8 bits acquired at 1Hz, and utilizing 3:1 lossless compression, resulting in a generation of data at the rate of 819.2 Kbps  
% % (https://thoracickey.com/digital-echocardiography/)

% The tasks are generated from a uniform distribution $\mathcal{U}[10\text{Kb},1\text{Mb} ]$ with computing density generated from a uniform distribution $\mathcal{U}[100,500 ]$.
We compare the results of our proposed solution to a benchmark of fixed cost for all tasks, set as the highest cost obtained across all of them, and a benchmark of a solution without bandwidth optimization. We 
\begin{figure}[ht]
    \centering
    \includegraphics[width=0.9\linewidth]{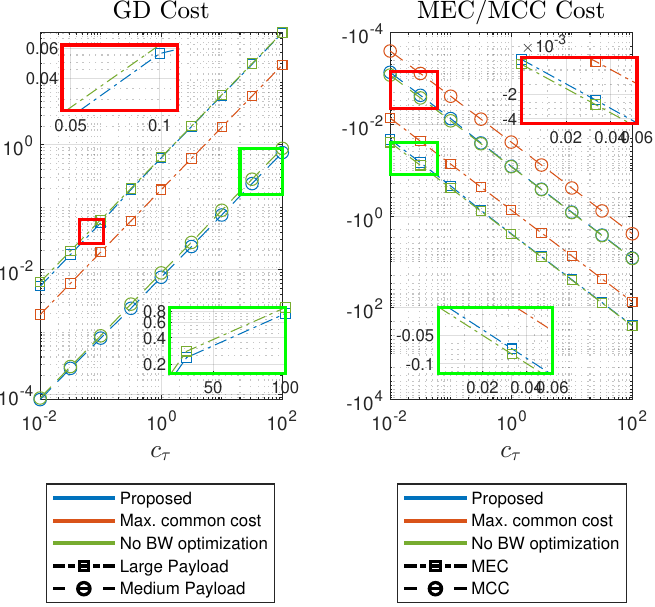}
    \caption{Per-tier cost over virtual delay cost $c_\tau$.}
    \label{fig:ctcb}
\end{figure}

In Fig~\ref{fig:ctcb}, we show the real cost (without the virtual delay cost) obtained by the GDs, the MEC and the MCC. Note that negative costs imply utility. As $c_\tau$ increases, the tasks are further incentivized to be offloaded to servers with higher capacity for faster computing of their tasks, thus the MEC and MCC can charge more for task computing. Nevertheless, this is only mostly true for the larger payload tasks, as low-payload tasks are computed locally with smaller delay than transmitting them. Furthermore, fixing the maximum cost dis-incentivizes offloading, reducing the utility at the MEC/MCC and the cost at the GDs, at the expense of faster task executions. On the other hand, not including the bandwidth optimization has the largest impact for smaller $c_\tau$ values, where it reduces the cost at the GDs at the expense of also reducing utility at the MEC/MCC.

\begin{figure}[ht]
    \centering
    \includegraphics[width=0.9\linewidth]{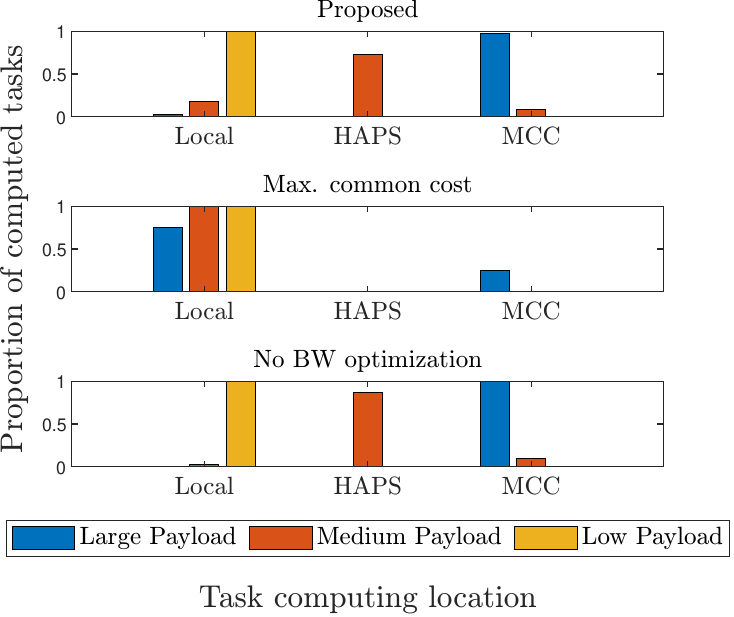}
    \caption{Computing location of tasks per task type.\vspace{-1em}}
    \label{fig:cifh}
\end{figure}

In Fig~\ref{fig:cifh} we show the location where tasks belonging to each of the three task types are computed. There is a clear tendency where higher-load tasks are computed at larger-resource nodes, where large payload tasks tend to be computed at the MCC, medium payload tasks tend to be computed at the HAPS MEC, and low payload tasks tend to be computed locally. Moreover, the proposed solution controls for remote infeasibilities and large costs in the bandwidth allocation problem, which causes more medium and large payload tasks to be computed locally, which prevents the remote servers from causing infeasibilities. Furthermore, the max. common cost benchmark is the one where the least amount of tasks are offloaded, due to the large costs for remote task processing.

\section{Conclusions}\label{sec:Conclusions}

We studied a hierarchical NTN edge-cloud task offloading architecture where each tier incurs costs when offloading tasks to a higher layer, while gaining profit from the tasks offloaded to it. 
% The availability of NTNs for remote healthcare applications provides computing resources for healthcare-related computing tasks, allowing for a faster computing, particularly for low-resource local devices. 
We proposed solutions for the optimal bandwidth utilization, as well as the per-task joint cost setting and offloading decision from the remote servers, to incentivize offloading while improving their utility. We observed that tasks are computed at different tiers, depending on their payload, and that our solution allocates tasks while avoiding remote infeasibilities. We also observed a tradeoff between costs incurred across tiers where our solution mostly increased the utility at the remote servers while increasing local costs due to offloading of locally infeasibile tasks. Future works may tackle the computing resource optimization as well as explicit local infeasibility management.

\section*{Acknowledgement}
This work received funding from Horizon Europe under the Marie Sklodowska Curie actions: ELIXIRION (GA 101120135).

\bibliographystyle{IEEEtran}
\bibliography{readme.bib}

\end{document}